\documentclass[runningheads,a4paper]{llncs}

\usepackage{amssymb}
\usepackage[tableposition=top]{caption}

\usepackage{float}
\usepackage[caption = false]{subfig}
\usepackage{graphicx}
\usepackage{algorithmic}
\usepackage{mdframed}
\usepackage[export]{adjustbox}
\usepackage{booktabs}
\usepackage{tabularx}
\usepackage[ruled,noline]{algorithm2e}

\usepackage[dvipsnames]{xcolor}
\usepackage{multirow}
\usepackage{colortbl}
\usepackage{enumitem}

\usepackage[misc,geometry]{ifsym}

\usepackage{url}
 \urldef{\mailsa}\path|{jmozucha, brossi}@mail.muni.cz
 |

\newcommand{\keywords}[1]{\par\addvspace\baselineskip
\noindent\keywordname\enspace\ignorespaces#1}

\usepackage{pgfplots}
\pgfplotsset{width=6cm,compat=1.9}

\authorrunning{Tamer Abdelaziz, Aya Sedky, Bruno Rossi, Mostafa-Sami M. Mostafa}

\begin{document}

\mainmatter  

\title{Identification and Assessment of Software Design Pattern Violations}
%
%
%
%
%

%
  \author{Tamer Abdelaziz\inst{1} \and Aya Sedky\inst{1} \and Bruno Rossi\inst{2} \and Mostafa-Sami M. Mostafa\inst{1}
  \textsuperscript{(\Letter)}}


 \institute{
 Faculty of Computers and Information, Helwan University, Cairo, Egypt\\
 \url{{tamer.a.yassen, aya, mostafa.sami} @fci.helwan.edu.eg}
 \and
 Faculty of Informatics, Masaryk University, Brno, Czech Republic\\
 \url{brossi@mail.muni.cz}
 }

%
%


\maketitle

%
%


\begin{abstract}
The validation of design pattern implementations to identify pattern violations has gained more relevance as part of re-engineering processes in order to preserve, extend, reuse software projects in rapid development environments. If design pattern implementations do not conform to their definitions, they are considered a violation. Software aging and the lack of experience of developers are the origins of design pattern violations. It is important to check the correctness of the design pattern implementations against some predefined characteristics to detect and to correct violations, thus, to reduce costs. Currently, several tools have been developed to detect design pattern instances, but there has been little work done in creating an automated tool to identify and validate design pattern violations. In this paper we propose a Design Pattern Violations Identification and Assessment (DPVIA) tool, which has the ability to identify software design pattern violations and report the conformance score of pattern instance implementations towards a set of predefined characteristics for any design pattern definition whether Gang of Four (GoF) design patterns by Gamma et al\cite{gof} or custom pattern by software developer. Moreover, we have verified the validity of the proposed DPVIA tool using two evaluation experiments and the results were manually checked. Finally, in order to assess the functionality of the proposed tool, DPVIA is evaluated with a dataset containing 5,679,964 Lines of Code (LoC)among 28,669 Java files in 15 open-source projects, with a large and small size of open-source projects that extensively and systematically employing design patterns, to determine design pattern violations and suggest refactoring solutions, thus keeping costs of software evolution. The results can be used by software architects to develop best practices while using design patterns. 

\end{abstract}

\newpage
\section{Introduction}

Software design patterns, as first formalized by Gamma et al.\cite{gof}, are general reusable solutions to commonly occurring design problems within a given context, that lead to the construction of well-structured, maintainable, and reusable software systems. In some Java applications, approximately 20\% of system classes participate in at least one GoF design pattern occurrence and those classes can represent from 15\% to 65\% of total classes \cite{designroles}\cite{designroles2}. In addition, program efficiency and productivity of development is increased 25-30 \% by applying correct patterns \cite{Riehle}, this leading to a considerable impact on the overall system design.

Design patterns are often mentioned as double-edged sword, applying the right pattern can be the system saviour \cite{Bautista} while applying a wrong one makes it disastrous and create many problems for system design. There are alternative design solution might produce better results than design pattern \cite{Ampatzoglou}. Alternative design solutions are functionally equivalent to design patterns and can be used when a design pattern is not the right solution for a specific design problem, they have been introduced for at least 13 out of 23 GoF design patterns in \cite{Ampatzoglou2}. 
Detection design patterns instances from source code is not too much difficult task with the help of many approaches of design pattern detection tools. However, a single design pattern has many different implementations according to system requirements, the intent would remain same and this modified form of pattern is known as variant \cite{Iyad}. So, it is very important to check the correctness of the applied design patterns to conform with their definition characteristics.


Lately, identification and assessment of design pattern violations has attracted the effort of the software engineering community. Design pattern violation occurs when design pattern implementations do not conform to their definitions. Software aging and the lack of experience of developers are two origins of design pattern violations. Whereas, software aging is caused by the failure of the product's owners to modify it to meet changing customer and business needs, and software application has been subject to a lot of changes e.g. modifications of functionalities, of methods, of classes, etc, these changes may degrade the overall system design \cite{Parnas}. It has been reported that the classes that participate in GoF design patterns change more often than the classes that do not participate in design pattern occurrences \cite{J.M.Bieman} \cite{M.Gatrell}. In addition, novice developers may not have enough knowledge to build design patterns correctly or simply may not aware of these good design pattern practices and use alternatives to solve well-known problems. Therefore, the usage of design patterns needs to be better supported and automated by a tool that would automatically provide information about the applied design pattern aspects.

The aim of this work is to introduce and describe an automated Design Pattern Violations Identification and Assessment (DPVIA) tool in order to detect violations of design patterns that occur in different project implementations, and to measure (conformance score) the degree of conformity of the design pattern implementations compared to their definition characteristics to provides a valuable insight on design pattern violations assessment. DPVIA tool helps the developer to determine design pattern rot and this form of violations destroys structural integrity of patterns and must be resolved.

The similarity score is calculated by many studies for different purposes such as Tsantalis DPD \cite{Tsantalis} that employs a graph similarity algorithm \cite{GraphAlgo}, which takes as input both the system and the pattern graph and calculates similarity scores between their vertices to detect design patterns candidates. In our approach, conformance score is calculated to detect design pattern violations and the score is reported to the developer in addition to violation details, and suggested solutions based on a predefined characteristics.

This paper is consist of five major sections. Section 1 describes motivational work and introduction of whole work. Section 2 focusing on current state of the art work related our approach. Section 3 discusses phases of the proposed DPVIA tool. Section 4 gives detail of our approach implementation, practical experiment and results. Finally, Section 5 is conclusion and provides useful insights for future work.

\section{Background and Related Work}

As the focus of this work lies on detect design pattern violations and their evaluation, we reviewed the early work of Izurieta and Bieman \cite{Izurieta} on type of design pattern violations called decay. Decay can involve the design patterns used to structure a system where classes that participate in design pattern realizations accumulate non pattern related code. Izurieta and Bieman investigated the evolution of design pattern implementations to comprehend how patterns decay and examined the extent to which software designs actually decay by studying the aging of design patterns in three successful object-oriented systems that include the entire code base of JRefactory, and added two additional open source systems \textemdash ArgoUML and eXist. The results indicate that pattern grime (non-pattern-related code) that builds up around design patterns is mostly due to increases in coupling and it is the main factor for the decay of software design patterns.

Pattern grime is defined as \textit{"degradation of the instance due to buildup of unrelated artifacts e.g., methods and attributes in pattern instances"} as a type of decay and divided the grime in to three categories \textemdash class, modular and organizational grime, and it has been pointed out as one recurrent reason for the decay of GoF pattern instances. Consequently, Izurieta in his doctoral dissertation \cite{Izurieta2} studied the accumulation of pattern decay and recognized another type of design decay called pattern rot. Furthermore, he noticed that this form of violations destroys structural integrity of design patterns. Pattern rot which is either a slow deterioration of software performance over time or its diminishing responsiveness that will eventually lead to software becoming faulty, unusable and in need of upgrade. Two distinct categories of design pattern decay were identified:
\begin{itemize}
    \item \textbf{Design Pattern Grime:} accumulation of unnecessary or unrelated software artifacts within the classes of a design pattern instance.
    \item \textbf{Design Pattern Rot:} violations of the structure or architecture of a design pattern.
\end{itemize}
Design pattern realizations can become a rot, when modifications of source code disrupt the structural or functional integrity of a design pattern. Design pattern rot due to failure to meet their responsibilities during pattern implementations, and thus represents a fault. In contrast with grime buildup does not break the structural integrity of a pattern but can reduce system testability and adaptability \cite{Izurieta1}.

Furthermore, Naouel Moha et al. \cite{Moha} defined a taxonomy of potential design pattern defects and conducted an empirical study to investigate their existence. The authors defined design pattern defects as errors occurring in the design of the software that come from the absence or the bad use of design patterns. The taxonomy includes the following four types of defects: \textit{An approximative or deformed design pattern} is a design pattern that has not been well conforming with GoF \cite{gof} definition but that is not erroneous. \textit{A Distorted or degraded design pattern} is a distorted form of a design motif which is harmful for the quality of the code. \textit{A Missing design pattern} is when a design is missing a needed design pattern. According to GoF \cite{gof}, missing patterns generates poor design. \textit{Excess design pattern} is the over use of design patterns in a software design. Later on, Izurieta cooperated with other researchers to obtain better comprehensions of patterns decay. Afterwards, Dale and Izurieta \cite{Dale}  proposed study on impacts of design patterns decay on quality of project.

Design patterns have been studied from various points of view by many authors. There has been little work done in creating an automated tool for validating instances of design patterns and identify violations that can be harmful to the design pattern instances realization and the overall system design. Primarily studies targeting design pattern validation such as Strasser et al. \cite{Strasser} focused on design patterns scoring where each candidate pattern is given a score, based on the resemblance with the design pattern definition. The author's proposed approach uses the Role-Based Metamodeling Language (RBML) \cite{Kim} in combination with PlantUML \footnote{PlantUML \url{http://plantuml.sourceforge.net/}} specification to calculate score of patterns conformance towards pattern definitions. The Role Based Metamodeling Language is a visually oriented language defined in terms of a specialization of the UML metamodel that is used to verify and specify generic or domain specific design patterns. The authors designed RBML-UML-Visualizer tool\footnote{ Strasser et al. automated tool is free and is available to download at \url{http://code.google.com/p/rbml-uml-visualizer/}} in order to inform developers when design patterns no longer conform to their original intended design. One of the drawbacks mentioned by the authors is that the algorithm only permits an UML object to be matched with an RBML model if the UML satisfies all of the RBML blocks requirements. Subsequently, some pattern instances cannot be evaluated without providing both RBML definitions and PlantUML specifications. In order to overcome those drawbacks the validation of design pattern instances should be done based on source code files directly without relying on RBML model or UML diagram.

In this paper, the proposed DPVIA tool validates instances of design patterns which are detected by the work of Diamantopoulos et al. \cite{Diamantopoulos} that proposed an open-source design pattern detection tool called DP-CoRe. Although some of software design pattern detection tools are effective for identifying several types of patterns, they have some drawbacks. For example, they require the source code to be compliable at least. As a result, developers cannot detect design pattern candidates without first resolving the source code issues and executing them correctly. Another drawback, most design pattern detection tools are designed as black box system that do not allow the developer any control over the detected patterns. Consequently, the proposed design pattern detection approach by Diamantopouloset al. is picked because it provides a solution for the mention drawbacks of other tools. DP-CoRe supports both the detection of 6 GoF patterns and offers the ability to add custom pattern definitions by the software developer. However, DP-CoRe depends on the latest compiler technology to enhance the detection of patterns instances in Java applications, DP-CoRe neither evaluates the conformance of pattern implementations towards pattern definitions nor focuses on measurement of their impact on code. The reason is that the tool is designed to detect pattern instances present in the source code, not to evaluate the correctness of their implementations. Consequently, we modified the open-source DP-CoRe tool to be a part of our automated tool DPVIA to identify design pattern violations and evaluate desired conformance scores by comparing pattern implementations to their definitions based on predefined characteristics.

\section{DPVIA: Software Design Pattern Violations Identification and Assessment Tool}
\begin{figure}
  \centering
  \includegraphics[width=\linewidth, frame]{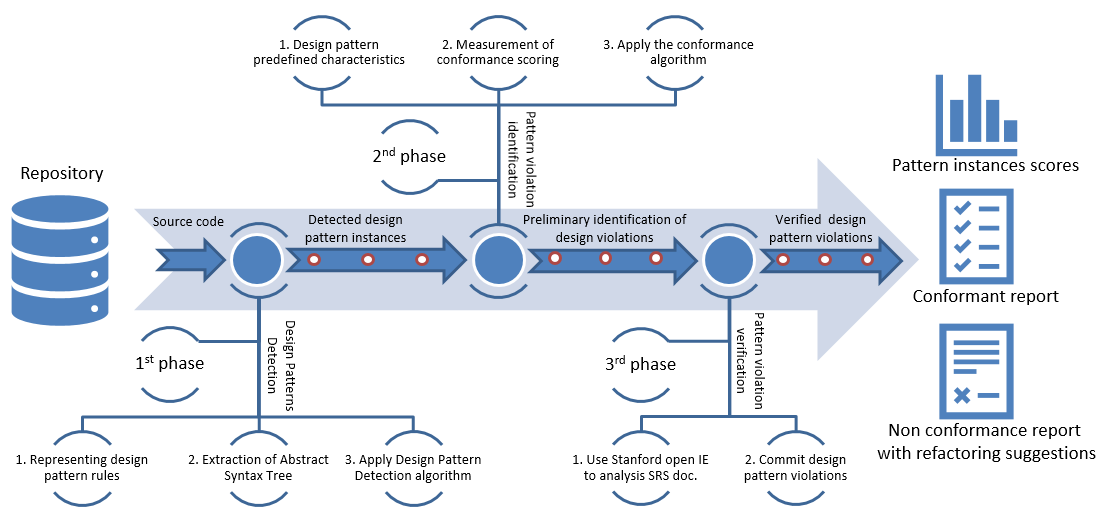}
  \caption{Phases of usage of the DPVIA tool}
  \label{fig:tool}
 \end{figure}

In this section, we describe the phases of the proposed tool, is shown in Figure \ref{fig:tool}. The first phase describes how DP-CoRe is integrated as part of DPVIA, and how design pattern detection approach, by Diamantopoulos et al. \cite{Diamantopoulos}, is working. The design pattern detection phase receives two inputs: the examined repository projects and the pattern abstraction \& connections rule files that could be modified by the software developer. The output is a list of detected pattern pattern instances, discussed in subsection \ref{Design Patterns Detection}. Thereafter, the tool calculates the conformance scores of the detected design pattern instances implementation versus their definitions in order to produce a preliminary identification of violations in the second phase, discussed in subsection \ref{Design pattern violation identification}. The last phase verifies the detected violations by examining relationships between entities participated in those violations according to system requirement specifications (SRS) document in format of IEEE template, this phase is implemented with help of the Stanford CoreNLP Natural Language Processing Toolkit \cite{StanfordNLP}. Consequently, the detected violation is considered a clear violation only if the relationship between violation entities is found in software business logic, discussed in subsection \ref{Verification of initial detected violations}. Finally, the proposed DPVIA tool reports the conformance scores of the detected pattern instances, and suggests a refactoring recommendations for the software developer to modify design pattern candidates and resolve their violations with minimum impact.

\subsection{Design Patterns Detection} \label{Design Patterns Detection}

We used \textbf{the proposed design pattern detection approach by Diamantopoulos et al.} \cite{Diamantopoulos}. We created rules of detection for 7 design patterns, at least two pattern for all categories: the creational patterns Simple Factory and Factory Method, the structural patterns Adapter and Decorator, and the behavioral patterns Observer, State and Strategy. The rules files is created based on two types of structural representations for source code and design patterns: the abstraction type of each class (e.g. Normal, Abstract, Interface, etc.) and the connection between two classes (e.g. inherits, calls, creates, has, uses and references) that is shown in Table \ref{table:characteristics}. The detection rules files and the examined repository projects are required as inputs for the pattern detection phase.

\begin{table}[h!]
\caption{Representing design pattern characteristics}
\label{table:characteristics}
\begin{tabular}{ p{3cm}p{9cm} } 

\hline 

Abstraction Type& Description\\
\hline
Normal& a non-abstracted class\\
Abstract& a Java abstract class\\
Interface& a Java interface\\
Abstracted& an abstract class or an interface\\
\hline

Connection Type& Description\\
\hline
A calls B& a method of class A calls a method of class B\\
A creates B& class A creates an object of type class B\\
A uses B& a method of class A returns an object of type B\\
A has B& class A has one or more objects of type B\\
A references B& a method of class A has as parameter an object of type B\\
A inherits B& class A inherits or implements class B\\
\hline
\end{tabular} 
\end{table}

The approach by Diamantopoulos et al. \cite{Diamantopoulos} starts with the extraction of Abstract Syntax Tree (AST) for each Java file using the Java Compiler Tree API and extract Java classes and relationships between them, Pattern candidates are then detected using the proposed detection algorithm that check all possible permutations of each class can be matched to the detection rules of pattern members. This is performed, as described in \cite{Diamantopoulos}, by recursively structured algorithm initialized with depth equal to 0. Iterating over the first class, it is checked whether its abstraction and its connections are the same with pattern member 0. If the matching is done, the detecting function is called recursively on the remaining classes except the already matched class and the depth is also incremented, else the recursive function stops. When all pattern members are matched successfully, then the Candidate is added to the detected pattern Candidates. An example output of \cite{Diamantopoulos} pattern detection approach is shown in Figure \ref{fig: output of detection phase}.
\begin{figure}
\begin{tabular}{ p{12cm} } 
\hline
\textit{ Candidate of Pattern Strategy: } \\
\textit{ A (Concrete Strategy): FlyRocketPowered } \\
\textit{ B (Strategy):          FlyBehavior } \\
\textit{ C (Concrete Context):  DecoyDuck }  \\
\textit{ D (Context):           Duck } \\
\hline 
\end{tabular}
    \caption{Example output of detection phase}
    \label{fig: output of detection phase}
\end{figure}

\subsection{Design pattern violation identification}
\label{Design pattern violation identification}

Upon having the list of detected design pattern candidates as output of previous subsection \ref{Design Patterns Detection}, the second phase of the proposed automated tool (DPVIA) is starting to evaluate the conformance of pattern candidate implementations compared to pattern definitions based on a predefined set of characteristics, in order to understand the violations that can occur when a design pattern is applied.

The design violations should be detected in early stages of evolution and based on their severity and overall pattern performance decide to keep, refactor or discard them. That is why the paper is centered on the identification of violations against design pattern definitions at first. Secondly focuses on measurement of their impact on source code and system design. Subsequently, the presence or absence of the abstraction of pattern candidate members and the connections among pattern members, if they are different from the predefined pattern characteristics, it is considered as a violation.

\subsubsection{Design pattern predefined characteristics:}

For each of the seven selected patterns, a set of predefined characteristics is created to address pattern specifications (e.g. abstraction of pattern classes and relationships characteristics). As well as, we arranged them with consideration of programming language specifications, which shaped the final concrete implementation. For purpose of obtaining characteristics comparable with patterns in real projects, which are implemented in one particular language have to be considered as well. We have decided to use the \textit{Java object oriented language} because there is fairly large amount of pattern definitions available and easily accessible in open source projects.


For instance, according to GoF \cite{gof} pattern definitions, Strategy predefined characteristics are described in Table \ref{table:Strategy Characteristics}.
\begin{table}[h!]
\caption{Strategy Design Pattern Predefined Characteristics}
\begin{tabular}
{|p{2cm}|p{2.5cm}|p{2.5cm}|p{3.1cm}|p{1.65cm}|  }

\hline
\multicolumn{5}{|l|}{}\\
\multicolumn{5}{|l|}{\textbf{Abstraction Predefined Characteristics }} \\
 \hline
Pattern Name & \multicolumn{2}{|l|}{Pattern Members (classes)} &Abstraction Type& Conforming \\
\hline
\multirow{4}{2cm}{\textbf{Strategy Pattern }} & \multicolumn{2}{|l|}{ConcreteStrategy} & Abstraction.Normal & required  \\ 
& \multicolumn{2}{|l|}{Strategy} & Abstraction.Interface & required \\ 
& \multicolumn{2}{|l|}{ConcreteContext} & Abstraction.Normal & optional \\ 
& \multicolumn{2}{|l|}{Context} & Abstraction.Normal & required \\ 
\hline
\multicolumn{5}{|l|}{}\\
\multicolumn{5}{|l|}{\textbf{Relationship Predefined Characteristics }} \\
 \hline
Relation& Relation From& Relation To& Connection Type & Conforming\\
 \hline
Inheritance &ConcreteStrategy& Strategy& Connection.inherits & required \\
Inheritance &ConcreteContext& Context& Connection.inherits & required \\ 
Composition &ConcreteContext& ConcreteStrategy& Connection.creates & required \\
Association &Context& Strategy& Connection.calls & required \\
Aggregation &Context& Strategy& Connection.has & required \\
Association &Context& Strategy& Connection.references & optional \\
Dependency &Context& Strategy& Connection.uses & optional \\
\hline
\end{tabular}

\label{table:Strategy Characteristics}
\end{table}
All predefined characteristics have same scoring weight, all differences are treated equally, we acknowledge that the scoring weights should be different from one characteristic to another and are determined by experts. The conforming of Strategy pattern predefined characteristics are:
\begin{itemize}
    \item Strategy (Required abstraction  conforming) 
    \begin{itemize}
        \item declares an interface common to all supported strategies. 
        \item Context uses this interface to call the strategy defined by a ConcreteStrategy (Required relationship).
    \end{itemize}
    \item ConcreteStrategy (Required abstraction conforming) 
    \begin{itemize}
        \item implements a concrete strategy using the Strategy interface (Required relationship).
    \end{itemize}
    \item Context (Required abstraction conforming) 
    \begin{itemize}
        \item is configured with a ConcreteStrategy object (Required relationship).
        \item maintains a reference to a Strategy object (Required relationship).
       \item may define an interface that lets Strategy access its data (Optional relationship).
    \end{itemize}
    \item ConcreteContext (Optional abstraction conforming).
    \begin{itemize}
        \item usually inherits the context and creates ConcreteStrategy object (Required relationships if Strategy pattern contains ConcreteContext as one of it's members).
    \end{itemize}
\end{itemize}
Absence of required characteristic is considered a clear violation, while absence of optional characteristic is not considered a violation. Nevertheless, presence of optional characteristics increases percentage of pattern member conforming score. Upon having design pattern predefined characteristics, the next step is to check the conformance of detected design pattern candidate implementations towards the predefined characteristics of design pattern.

\subsubsection{Measurement of conformance scoring:} \label{Measurement}
The similarity score is the measure of how much alike two data objects are. Similarity measure in a programming context is a distance with dimensions representing features of the objects. If this distance is small, it will be the high degree of similarity where large distance will be the low degree of similarity. Similarity are measured in the range 0 to 1 [0,1]. Two main consideration about similarity:
\begin{itemize}
    \item Similarity = 1  if X = Y         (Where X, Y are two objects)
    \item Similarity = 0  if X $ \neq $  Y 
\end{itemize}
The purpose of measurement is obtaining a conformance scores between the predefined characteristics of pattern definitions and their implementations in source code. For all detected pattern candidate members, \textbf{our proposed conformance algorithm}, is shown in Figure \ref{fig:conformance algorithm},
\begin{figure}
\begin{algorithm}[H]
\SetAlgoLined
\KwResult{$PercentageOfPatternMember_{Score}$ }
 \textbf{CheckConformance}( PatternCharacteristics C, PatternCandidateMember M)
 $Scores Matrix \gets null$, $i \gets 0$\;
 \While{characteristic in C}{
 \If{C.characteristic \textbf{is} AbstractionType}{
 	\uIf{C.getAbstraction() \textbf{and} M.getAbstraction()}{
    	$Scores [i] \gets [1, 1]$
    }
    \uElseIf{C.getAbstraction() \textbf{and} ! M.getAbstraction()}{
    	$Scores [i] \gets [1, 0]$
    }
    \uElseIf{! C.getAbstraction() \textbf{and} M.getAbstraction()}{
    	$Scores [i] \gets [0, 1]$
    }
 }
 \If{C.characteristic \textbf{is} ConnectionType}{
 	\uIf{C.getConnection() \textbf{and} M.getConnection()}{
    	$Scores [i] \gets [1, 1]$
    }
    \uElseIf{C.getConnection() \textbf{and} ! M.getConnection()}{
    	$Scores [i] \gets [1, 0]$
    }
 }
 \textbf{print} violation details \textbf{and} suggested solution \newline
 $i \gets i + 1$
 }
 \textbf{return} \newline
 $PercentageOfPatternMember_{Score} \gets ( 1 - \frac{1}{ScoresSize}  \sum_{k=1}^{ScoresSize}  Scores1_{st}vector[k] \otimes Scores2_{nd}vector[k] ) * 100$
 
\caption{The proposed conformance algorithm}
\end{algorithm}
\caption{The proposed conformance algorithm}
\label{fig:conformance algorithm}
\end{figure}
receives two inputs as parameters for \textit{CheckConformance} function: first input is one of the pattern classes that participates in the predefined characteristics (e.g. class ConcreteStrategy, Strategy, ConcreteContext or Context characteristics, as shown in Figure \ref{table:Strategy Characteristics}), and the second input is the corresponding pattern candidate member (e.g.the corresponding class of the pattern implementations). At first, the algorithm is initialized with empty scores matrix then iterating over all possible characteristics, check characteristic type (e.g. abstraction or connection) then compare it with the corresponding pattern candidate member and add value to similarity scores matrix according to fulfilled condition. While doing so, we noticed that only the limited scenarios depicted in Table \ref{table:ComparingScenariosTable} would apply.
\begin{table}[h!]
\caption{Design pattern characteristics comparing scenarios}
\begin{tabular}
{p{1.5cm}p{1.5cm}p{6.5cm}{r}}

\hline

Predefined characteristic& Candidate member implementation& Explanation& Representation \\
 \hline
True & True & The characteristic is present in predefined characteristics of pattern definition as well as in the implementation of pattern candidate member source code & [1, 1]\\
&&&\\
True & False & The characteristic is present in predefined characteristics of pattern definition but is not in the implementation of pattern candidate member source code & [1, 0]\\
&&&\\
False & True & The characteristic is not present in predefined characteristics of pattern definition but can be found in the implementation of pattern candidate member source code & [0, 1]\\
&&&\\
False & False & The characteristic is not present in predefined characteristics of pattern definition and neither is in the implementation of pattern candidate member source code & [0, 0]\\
\hline
\end{tabular}

\label{table:ComparingScenariosTable}
\end{table}

Similarity scoring is represented by a matrix of two vectors, where the first vector refer to absence or presence \textit{(0 or 1)} of a characteristic in the pattern definition characteristics while second vector serves the same purpose only for the pattern candidate member implementation. Consequently, for each characteristic in  the pattern definition characteristics has a complete satisfaction with the corresponding implementation of pattern candidate member of source code, the value \textit{[1, 1])} will be added in the scoring matrix. While the characteristic is present in definition but is absent in pattern member indicate inconsistency and is considered a clear violation by adding value \textit{[1, 0]} to scores matrix. However, the absence of a particular definition characteristic and its presence in pattern member is not necessarily to be a violation and gives an equal probability for identification of violation or normal artifact. Therefore, this situation is considered a violation for abstraction characteristic types only, because for every pattern candidate member in source code has only one abstraction characteristic type (class type), and if it does not match the corresponding pattern definition abstraction, it must be defined as violation by adding value \textit{[0, 1]} to scores matrix. The awareness of absence of characteristic from pattern member and also its non existence in definition characteristics, does not add  anything about similarity score, so that double negative value \textit{[0, 0]} is recognized as non-valuable information for similarity measure with in this work. Finally, we use the most straight forward way to measure the similarity between two vectors of the similarity matrix and return the conformance score by formula (\ref{eq}):
\begin{equation} \label{eq}
PercentageOfPatternMember_{Score} = ( 1 - \frac{1}{N}  \sum_{i=1}^{N}  C_{i} \otimes M_{i} ) * 100
\end{equation} 
Where:\newline
\textit{PercentageOfPatternMember{\tiny Score}} is the conformance score percentage, \textit{N} is the similarity matrix rows (size of characteristics), \textit{C{\tiny i}} is the pattern definition characteristic binary value representing by the \textit{$1_{st}$ vector} of similarity score matrix, and \textit{M{\tiny i}} is the pattern candidate member binary value representing by the \textit{$2_{nd}$ vector} of similarity score matrix.

\subsubsection{An illustration of design pattern violation identification: }
\begin{figure}
  \centering
  \includegraphics[width=\linewidth, frame]{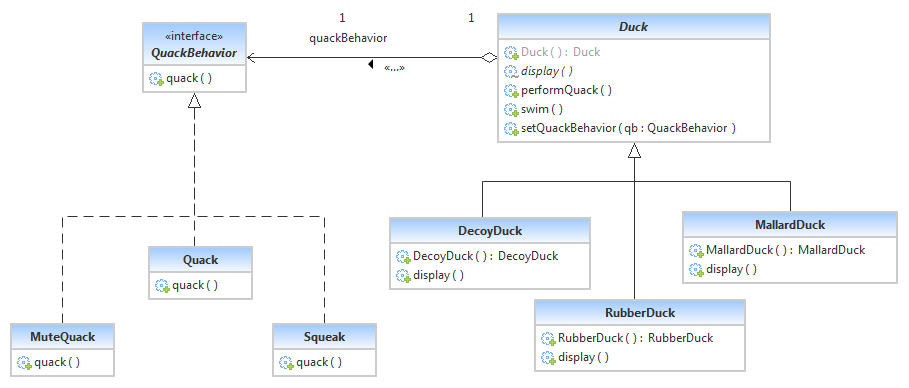}
  \caption{Strategy candidate instances UML class diagram}
  \label{fig:Strategy candidate instances UML class diagram}
\end{figure}

For example, in Strategy design pattern, consider the following 3 Strategy candidate instances, shown in Table \ref{table:Strategy Candidate Instances} and visualized in Figure \ref{fig:Strategy candidate instances UML class diagram}, are detected by approach by Diamantopoulos et al. \cite{Diamantopoulos} in the first phase of DPVIA tool.
\begin{table}[h!]
\caption{Strategy Candidate Instances}
\begin{tabular}
{p{3cm}p{3cm}p{3cm}p{2.8cm} }

\hline
\textbf{Pattern Members}& \textbf{Candidate \#1}& \textbf{Candidate \#2}& \textbf{Candidate \#3}\\
 \hline
ConcreteStrategy &Quack& Squeak& MuteQuack\\
Strategy&QuackBehavior& QuackBehavior& QuackBehavior\\ 
ConcreteContext &MallardDuck& RubberDuck& DecoyDuck\\
Context &Duck& Duck& Duck\\
\hline
\end{tabular}

\label{table:Strategy Candidate Instances}
\end{table}
Strategy pattern, in this example, represents a family of \textit{Quack Behaviour strategies}, encapsulate each one, and make them interchangeable. Strategy lets the algorithm vary independently from clients that use it. Each candidate has 4 members:
\begin{itemize}
    \item ConcreteStrategy
    \item Strategy
    \item ConcreteContext
    \item Context
\end{itemize}
As shown in Table \ref{table:Strategy Candidate Instances}, class \textit{Duck} represents Context member of the three Strategy candidates. In this example, we show how our proposed approach measures the conformance of \textit{Duck} class towards \textit{Context} member of Strategy predefined characteristics described in Table \ref{table:Strategy Characteristics}, using the proposed conformance algorithm showed in Figure \ref{fig:conformance algorithm}, as following in Table \ref{table:Example of Conforming}.
\begin{table}[h!]
\caption{Measurement of conformance scoring example}
\begin{tabular}
{p{6.5cm}p{1.5cm}p{1.5cm}{r}}

\hline

Predefined Characteristic & Pattern member (Context)& Candidate member (Duck)& Scores Matrix \\
 \hline
Abstraction.Normal (required) & True & True  & [1, 1]\\
&&&\\
Connection.calls (required) to Strategy & True & False  & [1, 0]\\
&&&\\
Connection.has (required) to Strategy & True & True  & [1, 1]\\
&&&\\
Connection.references (optional) to Strategy & True & True  & [1, 1]\\
&&&\\
Connection.uses (optional) to Strategy & True & True  & [1, 1]\\
\hline
\end{tabular}
\label{table:Example of Conforming}
\end{table}

The proposed conformance score formula (\ref{eq}), in the proposed conformance algorithm, uses the Hamming Distance algorithm, one of the most popular similarity distance measures, that denote the difference between two binary vectors of equal length. It is the number of positions at which the corresponding symbols are different \cite{Hamming}. In this example, the Hamming Distance of two binary vectors of the scores matrix that is shown in last column of Table \ref{table:Example of Conforming} whereas first vector: \textit{[1, 1, 1, 1, 1]} and second vector: \textit{[1, 0, 1, 1, 1]} is calculated as following steps:
\begin{itemize}

    \item Step 1
    Ensure the two vectors are of equal length. The Hamming distance can only be calculated between two vectors of equal length. 

    \item Step 2
    Compare the first two bits of both vectors. If they are the same, record a "0" for that bit. If they are different, record a "1" for that bit. In this example, the first bit of both vectors is "1," so record a "0" for the first bit.

    \item Step 3
    Compare each bit in succession and record either "1" or "0" as appropriate. For vector 1: \textit{[1, 1, 1, 1, 1]} and vector 2: \textit{[1, 0, 1, 1, 1]}, the record \textit{[0, 1, 0, 0, 0]} is obtained.

    \item Step 4
    Add all the ones and zeros in the record together to obtain the Hamming distance. \textit{Hamming distance = 0 + 1 + 0 + 0 + 0  = 1}.
\end{itemize}
The two binary vectors have \textit{1} different bit, this is what constitutes the cornerstone of formula (\ref{eq}). So the percentage of pattern member conformance score (class \textit{Duck}) = (1 - (1/5) ) * 100 = 80 \%. Because of class \textit{Duck} implementation missed calling \textit{ quackBehavior.quack();} to perform \textit{quack} behavior, it is considered a clear violation. Assume that class \textit{Duck} does not define an interface that lets Strategy access its data (Optional relationships), this absence of optional connections is not considered a violation but the conformance score will be (1 - 1/3) * 100 = 66.66 \%.

After measuring the conformance scores for all pattern candidate members, the average is calculated for the pattern candidate as a whole and the score is reported to the developer in addition to in order to produce a preliminary identification of violation details, and suggested solutions based on previously defined characteristics. The proposed approach suggests refactoring for all violations. For instance, the missing of call connection in class \textit{Duck} to perform \textit{quack} behavior that detected as violation could be solved as following: \newline \newline
\textit{Recommendation - Class( Duck ) should calls (invoke function quack) of class QuackBehavior}.\newline
\newline
Such suggestions help developers to resolve violations and providing a valuable insight on "health" of system under study and possible existence of violations within its source code. In order to distinguish between code related to design pattern realization and code that is harmful causes a decay of system design.

\subsection{Verification of the initial detected violations} 
\label{Verification of initial detected violations}

 Finally, the last phase of DPVIA tool verifies the detected violations by examining relationships between entities participated in those violations based on the presence / absence of relationship scenarios between those entities, in system requirement specifications (SRS) document in format of IEEE template. In order to take business logic constrains into consideration before accounting those detected violations in the total conformance score.
 
 In our proposed approach, the Natural Language Processing Toolkit \cite{StanfordNLP} is required to extract the entities relationship scenarios of the project under study. We integrated the proposed DPVIA tool with a Java implementation of Stanford Open Information Extraction (open IE) as described in the paper of Gabor Angeli et al. \cite{StanfordOperIE}. Open IE refers to the extraction of relation tuples, typically binary relations, from plain text. The central difference is that the schema for these relations does not need to be specified in advance; typically the relation name is just the text linking two arguments.

 The open IE first splits each sentence into a set of entailed clauses. Each clause is then maximally shortened, producing a set of entailed shorter sentence fragments. These fragments are then segmented into OpenIE triples, and output by the system. An illustration of the process is given for an example sentence below in Figure \ref{fig:Open IE example}:
 
 \textit{"Employee opens the control panel, view all complaints and solve client problems"}
 
 \begin{figure}[h]
      \centering
      \includegraphics[width=\textwidth,frame]{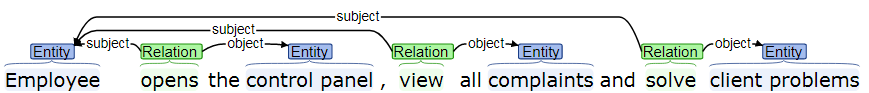}
     \caption{Stanford OpenIE example}
      \label{fig:Open IE example}
\end{figure}

All extracted relationships between subject and object entities are stored in Java list collection. Consequently, the detected violation is considered a clear violation only if the relationship between violation entities is found in the stored Java list collection, or discard the detected violation, if there is no relationships in business logic between violation entities.

Finally, the proposed approach reports the pattern instance scoring with refactoring suggestion to modify Java application with minimum impact. In order to guide the developer to enhance and extend software applications by supporting an assessment score of current source code implementations and recommendation to solve design violations.

\section{Implementation, Practical Experiment and Results} \label{Evaluation}

Practical experiments are done to study, using the proposed approach, \textbf{how would design patterns be applied in real environment of open source projects in order to assess the implementations of software design patterns, detect design pattern violations, and offer recommendations for resolve those violations.}

\subsection{Implementation of the proposed approach}

Our proposed approach\footnote{The automated tool DPVIA is free and is available to download at \url{https://github.com/TamerAbdElaziz/DPVIA}} DPVIA is implemented in Java programming language, we have decided to use the Java object oriented language because it is one of mainstream programming languages nowadays, thus there is fairly large amount of pattern definitions available. Consequently, finding open source projects with easily accessible source codes is not an issue.

The DPVIA tool offering a \textit{Command Line Interface} (CLI) to obtain the identification of design pattern violations in Java repository projects then reports the conformance scores for all pattern candidates and violations details with recommended solutions. In addition, It produces graphs indicating the percentage of violation that has been committed. 

The automated tool is free and available to download it from Git or checkout with SVN using the web URL:\textit{  \url{https://github.com/TamerAbdElaziz/DPVIA.git}}, then unzip the downloaded file. There will be two folders named \textit{"pattern"} and \textit{"Repository"}, as well executable Jar file named \textit{"dpvia"}, then follow the following instructions:
\begin{itemize}
    \item The DPVIA is able to detect pattern violations successfully of 7 design patterns as mentioned before, it offers the ability to define custom patterns by the developer. Any design pattern characteristics could be defined and added to folder that named \textit{"pattern"}.
    
    \item The developer is able to set any Java project source code files on the folder called \textit{"Repository"}. As well, many projects can be examined at one time.
    
    \item Run in batch (command line) mode of Jar file which called \textit{dpvia}, and execute using command: \textit{java -jar dpvia.jar}
\end{itemize}
The inputs to DPVIA tool is any set of Java projects source code. In the other hand, the final output is formatted as comma-separated values (CSV) file stores tabular data (numbers and text) in plain text about each design pattern member assessment and recommendation of solution if there is violations. In addition to CSV file, the assessment is visualized using Bar Chart and the recommendations is written in word document.

\subsection{Practical experiments}

DPVIA is evaluated in Java project of \textit{Head First Design Patterns Book code} \footnote{Head First Design Patterns Book code is free and available to download it from Headfirstlabs website using the web URL:\textit{ \url{http://www.headfirstlabs.com/books/hfdp/HeadFirstDesignPatterns_code102507.zip}.}} which provides an interesting example project that has a proper implementations of well-known design pattern patterns (e.g. Simple Factory, Factory Method, Adapter, Decorator, Observer, State and Strategy). Note, we have modified some instances of this project to make them contain violations. The validation of the proposed tool (DPVIA) using two evaluation experiments:

\subsubsection{The first practical experiment:}
\label{The first practical experiment}
Integration of our approach with DP-CoRe tool (in DPVIA first phase) has succeeded in determining all design pattern candidates with accuracy 70.73\% of the detection algorithm where 24 of pattern candidates were detected incorrectly (false positive 29.26\%) while 58 of pattern candidates were detected correctly. Moreover, by reviewing the source code manually, we found the total number of the correct pattern candidates in source code is 58 candidates, so no candidates were missed without detection, but some of the detected instances are not fully representative of design patterns. Pattern detection algorithm by DP-CoRe achieved 70.73\% precision and 100\% recall. 

Then DPVIA (in DPVIA second phase) has measured the conformance score for each detected pattern candidate in order to identify pattern violations and report the conformance scores average, satisfied and violated instances of the examined project, the results are shown in Table \ref{table:result2}.
\begin{table}[h]
\caption{Validating the proposed approach over \textit{Head First Design Patterns Book code} project}
\label{table:result2}

\begin{tabular}{|p{1.5cm}| p{1.7cm} p{1.7cm}| p{1.9cm} p{1.5cm} p{1.5cm} p{1.5cm}|}

\hline

& \multicolumn{2}{|c|}{Design Patterns Detection}& \multicolumn{4}{|c|}{Design pattern violation identification}\\

\hline  
 
Pattern name&\#Instances  & \#Incorrect Instances detection  &Conformance Score \%& \#Satisfied Instances&  \#Violated Instances & \#Incorrect Instances Scoring\\
\hline 
 
 Adapter & 2 & 0 & 100\% & 2 & 0 & 0 \\
 Decorator & 16 & 0 & 96.2\% & 8 & 8 & 0  \\
 FactoryM & 16 & 0 & 100\% & 16 & 0 & 0  \\
 SFactory & 4 & 0 & 100\% & 4 & 0  & 0  \\
 Observer & 4 & 0 & 92.5\% & 2 & 2  & 0  \\
 State & 5 & 0 & 96\% & 3 & 2 &  0  \\
 Strategy & 35 & 24 & 93.9\% & 10 & 25 & 24 \\
 
\hline
Total & 82 & 24 &  & 45 & 37  & 24 \\
\hline 
\% of Total &   & 29.26\% &  & 54.87\% & 45.12\%  & 29.26\% \\
\hline 
\end{tabular} 
\end{table}
In the fourth column shows the average of conformance scoring for each pattern in the range of 92.5\% to 100\%. The conformance scoring was verified manually by reviewing the source code of the satisfied and violated instances, we found 24 instances were identified as violated instances incorrectly (false positive 29.26\% of the proposed conformance scoring algorithm). The proposed conformance algorithm achieved 70.73\% precision and 100\% recall.

Consequently, the conformance algorithm has false disclosure due to the measurement of conformance score of some pattern instances were detected in the detection phase incorrectly and the reliance only on predetermined characteristics of each design pattern while it should not be considered a violation according to business logic and software requirements. For this reason, we suggested the verification phase for the detected violations. Verification phase could be done by software developers but it needs a lot of time and effort. If the relationships between system entities in the SRS document are presented to the software developer, it will be easy to approve or discard the violations based on the presence or absence relationships between violation members or perform the verification phase automatically.

\begin{figure}[h]
      \centering
            \includegraphics[width=\textwidth,frame]{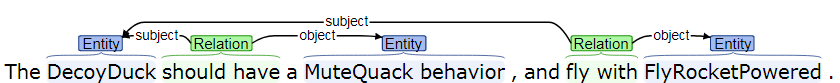}
     \caption{Stanford Open Information Extraction of relationships between entities
     }
      \label{fig:Open IE}
\end{figure}

The proposed tool (DPVIA) is integrated with Stanford Open Information Extraction (open IE) \cite{StanfordOperIE} that extracts open-domain relation triples, representing a subject, a relation, and the object of the relation from plain text. Open IE can be accessed through the Stanford CoreNLP API\footnote{Stanford CoreNLP \url{https://stanfordnlp.github.io/CoreNLP/}} through the standard annotation pipeline to extract the relations between violation members from SRS plain text. An illustration of the process is given for an example sentence below which is written in SRS document and represented in Figure \ref{fig:Open IE}:\newline
\begin{center}
    \textit{"The DecoyDuck should have a MuteQuack behavior, and fly with FlyRocketPowered"} \newline
\end{center}
According to the extraction of relations between entities, the entity \textit{DecoyDuck} has only two relations with \textit{MuteQuack} behavior and \textit{FlyRocketPower}. However, during pattern detection and violation identification, \textit{DecoyDuck} entity participates as member class in 7 detected Strategy instances where 2 instances conformed the predefined characteristics while other 5 instances did not. The five violated instances, \#4, \#9, \#14, \#24, \#29, have a missing connection from class(\textit{DecoyDuck}) to class (\textit{Squeak}), class (\textit{FakeQuack}), class (\textit{Quack}), class (\textit{FlyWithWings}) or class (\textit{FlyRocketPowered}) respectively. So that, the violations of Strategy instances \#4, \#9, \#14, \#24 were discarded due to the absence of relationships between violation members in the result of open IE relations extraction. The only instance \#29 is considered as violation where \textit{DecoyDuck}, in source code, flies with another flying behavior and does not fly with \textit{FlyRocketPowered} behavior as required. The result of instance \#29, as shown in the Figure \ref{fig:pattern29}, shows how DPVIA tool is able to detect design pattern violations and recommend a suitable refactoring solutions.
\begin{figure}
\begin{tabular}{ p{12cm} } 
\hline

\textbf{\textit{Candidate of Pattern Strategy (29): }} \\
\textit{A(Concrete Strategy): FlyRocketPowered } \\
\textit{B(Strategy): FlyBehavior } \\
\textit{C(Concrete Context): DecoyDuck } \\
\textit{D(Context): Duck } \\
\\

\textbf{\textit{Design pattern violation identification:}}

\\

\textit{FlyRocketPowered (Evaluation : 100.0 \%) } \\
\\
\textit{FlyBehavior (Evaluation : 100.0 \% )} \\
\\
\textit{DecoyDuck (	Evaluation : 66.0 \% )} \\
\textit{Recommendation: Class( DecoyDuck ) should creates new object of class : FlyRocketPowered } \\
\textit{Approved: This violation has to be solved according to the relationship between ( decoyduck ) and ( flyrocketpowered ) in SRS document. } \\
\\
\textit{Duck (Evaluation : 100.0 \% )} \\

\\
\textbf{\textit{Total score : 91.5 \% }} \\

\hline
\end{tabular}
    \caption{Example Output of DPVIA}
    \label{fig:pattern29}
\end{figure}

One of the most important results of the verification phase is the reduction of false positive instances scoring and is changed to be more accurate for the proposed conformance scoring algorithm. Currently, the verification phase of pattern violations works successfully only if the source code classes have the same system entity names in the SRS document. This issue could be solved by applying more accurate requirements analysis techniques.

\subsubsection{The second practical experiment:}
\label{The second practical experiment}

We repeated the same previous experiment with different settings of design pattern detection algorithm. Tsantalis DPD tool, uses similarity algorithms, is used to detect design pattern instances instead of Diamantopoulos et al. \cite{Diamantopoulos} algorithm used in previous experiment, then apply the same conformance scoring algorithm and running over the same project of \textit{Head First Design Patterns Book code}.


We got a set of detected pattern instances by Tsantalis DPD tool, and wrote the instances in a file named \textit{"PatternsDetectedByOtherTools.txt"} in the main path of DPVIA tool. The pattern instances are written in the following formats shown in Figure \ref{fig:otherToolFormats}. In addition, using these formats allows any developer has detected the pattern classes by other detection approaches to measure the conformance score easily and detect pattern violations.
\begin{figure}
\begin{tabular}{ p{12cm} } 
\hline

\textit{Decorator Espresso A Concrete Component } \\
\textit{Decorator Beverage B Component } \\
\textit{Decorator Soy C Concrete Decorator } \\
\textit{Decorator CondimentDecorator D Decorator } \\
\textit{End } \\
\textit{FactoryMethod NYStyleClamPizza A Concrete Product } \\
\textit{FactoryMethod Pizza C Adapter B Product } \\
\textit{FactoryMethod NYPizzaStore C Concrete Creator } \\
\textit{FactoryMethod PizzaStore D Creator } \\
\textit{End } \\
.\\
.\\
.\\
\textit{End } \\
\hline
\end{tabular}
    \caption{Formats of pattern instances detected by any detection tool}
    \label{fig:otherToolFormats}
\end{figure}

As it is obvious in Table \ref{table:result3}, Tsantalis DPD tool is totally missed detection of Simple Factory and Strategy pattern candidates and 15 of pattern candidates were detected incorrectly (false positive 65.21\%) while 8 candidates were detected correctly. As noted by the first experience, the total number of the correct pattern candidates in source code is 58 candidates, so 50 candidates were missed without detection (false negative 86.20\%). Pattern detection algorithm by Tsantalis DPD achieved 34.78\% precision and 13.79\% recall. 

Then DPVIA (in DPVIA second phase) has measured the conformance score for each detected pattern candidate. Note that pattern instances that are detected incorrectly by Tsantalis DPD might mislead the proposed conformance scoring algorithm (Fig. \ref{fig:conformance algorithm}) to assess of the violations correctly.
\begin{table}[h]
\caption{Validating the conformance algorithm integrated with Tsantalis DPD over \textit{Head First Design Patterns Book code} project}
\label{table:result3}
\begin{tabular}{|p{1.5cm}| p{1.7cm} p{1.7cm}| p{1.9cm} p{1.5cm} p{1.5cm} p{1.5cm}|}

\hline

& \multicolumn{2}{|c|}{Design Patterns Detection}& \multicolumn{4}{|c|}{Design pattern violation identification}\\

\hline  
 
Pattern name&\#Instances  & \#Incorrect Instances detection  &Conformance Score \%& \#Satisfied Instances&  \#Violated Instances & \#Incorrect Instances Scoring\\
\hline 
 
 Adapter & 10 & 8 & 69\% & 0 & 10 & 2 \\
 Decorator & 2 & 0 & 90\% & 0 & 2 & 2  \\
 FactoryM & 3 & 1 & 66.7\% & 2 & 1 & 0  \\
 SFactory & - & - & - & - & -  & -  \\
 Observer & 1 & 0 & 87.5\% & 0 & 1  & 1  \\
 State & 7 & 6 & 83\% & 0 & 7 &  1  \\
 Strategy & - & - & - & - & - & - \\
 
\hline
Total & 23 & 15 &  & 2 & 21  & 6 \\
\hline 
\% of Total &   & 65.21\% &  & 8.69\% & 91.30\%  & 26.08\% \\
\hline 
\end{tabular} \end{table}
In the fourth column in Table \ref{table:result3} shows the average of conformance scoring for each pattern. The Simple Factory and Strategy pattern have not had any conformance scoring because they were not discovered using Tsantalis DPD. Other design patterns are in rang of conformance scoring between 66.7\% to 90\% when they are compared to the predefined characteristics. The conformance scoring was verified manually by reviewing the source code of the satisfied and violated instances, we found 6 instances were identified as violated instances incorrectly (false positive 26.08\% of the proposed conformance scoring algorithm). The proposed conformance algorithm achieved 73.91\% precision and 100\% recall.

\subsection{Discussion and Results}
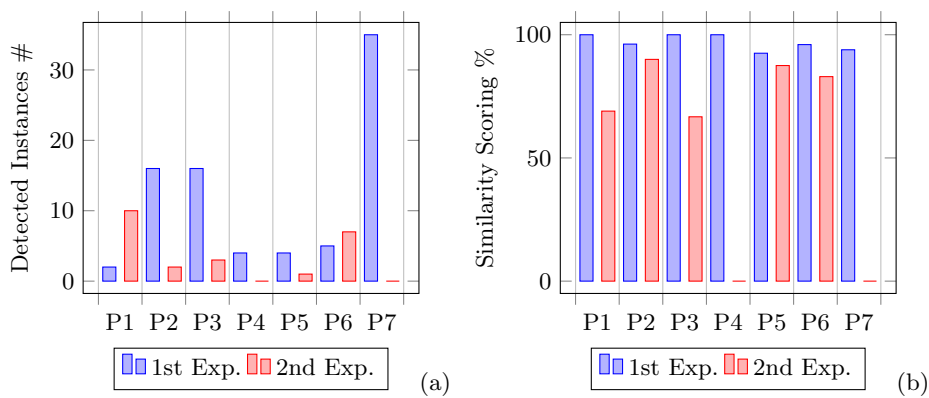
\begin{figure}
\begin{tabular}{p{6cm} p{6cm}}
\begin{tikzpicture}
\begin{axis}[
    x tick label style={
		/pgf/number format/1000 sep=},
        symbolic x coords={P1, P2, P3, P4, P5, P6, P7, P8},
        xtick=data,
	ylabel=Detected Instances \#,
	enlargelimits=0.05,
	legend style={at={(0.5,-0.2)},
	anchor=north,legend columns=0},
	ybar interval=.6,
]
\addplot 
		coordinates {(P1,2) (P2,16) (P3,16) (P4,4) (P5,4) (P6,5) (P7,35) (P8,0)};
\addplot 
	coordinates {(P1,10) (P2,2) (P3,3) (P4,0) (P5,1) (P6,7) (P7,0) (P8,0)};
\legend{1st Exp.,2nd Exp.}
\end{axis}
\end{tikzpicture}(a)
&
\begin{tikzpicture}
\begin{axis}[
    x tick label style={
		/pgf/number format/1000 sep=},
        symbolic x coords={P1, P2, P3, P4, P5, P6, P7, P8},
        xtick=data,
	ylabel=Similarity Scoring \%,
	enlargelimits=0.05,
	legend style={at={(0.5,-0.2)},
	anchor=north,legend columns=0},
	ybar interval=.6,
]
\addplot 
	coordinates {(P1,100) (P2,96.2) (P3,100) (P4,100) (P5,92.5) (P6,96) (P7,93.9) (P8,80)};
\addplot 
	coordinates {(P1,69) (P2,90) (P3,66.7) (P4,0) (P5,87.5) (P6,83) (P7,0) (P8,80)};
\legend{1st Exp.,2nd Exp.}
\end{axis}
\end{tikzpicture}(b)
\\ 
\end{tabular} 
\caption{Comparison between the two evaluation experiments (a) number of detected instances (b) Similarity scoring percentage.}
\label{fig:2Exp Comparison}
\end{figure}

The results for the two experiments are shown in Figure \ref{fig:2Exp Comparison}, where P1, P2, P3, P4, P5, P6, and P7 refer to enumerating patterns Adapter, Decorator, Factory Method, Simple Factory, Observer, State, and Strategy respectively. In Figure \ref{fig:2Exp Comparison} (a), there are large deviations between the detected patterns of the two experiments for the same project of \textit{Head First Design Patterns Book code}, which are mostly due to the detection algorithm of each experiment. Whereas, the detection algorithm by Diamantopoulos  et  al. \cite{Diamantopoulos} used in our proposed tool (DPVIA), in the first experiment, allowing developers the flexibility to specify a set of rules to detect any pattern, in contrast to that, the detection algorithm by Tsantalis DPD tool \cite{Tsantalis}, in the second experiment, uses similarity algorithms to detect patterns as a black box that do not allow the developer any control over the detected patterns. On other hand, in Figure\ref{fig:2Exp Comparison} (b), illustration of similarity scoring percentage of the two experiments. 

As already noted, \textbf{\textit{the conformance scoring correctness of pattern instances rely on the correct detection of those pattern instances}}, the interesting aspect of this finding is showing the importance of pattern detection algorithm in evaluation of design pattern violations. Also we observed, \textbf{\textit{DPVIA tool is quite effective for identifying design pattern violations}}, due to the flexibility to use any pattern detection rules as well as determine a set of characteristics that is used in measurement of conformance scores. Furthermore, \textbf{\textit{concerning execution time, our proposed tool is quite efficient}} whereas the identification and assessment of 58 design pattern instances in \textit{Head First Design Patterns Book code} project that contains 2,063 Lines of Code (LoC), required almost 2.5 seconds.

In order to assess the functionality of the tool on any open source project, \textbf{\textit{DPVIA is evaluated with a dataset containing 5,679,964 (LoC) Lines of Code among 28,669 Java files in 15 open-source projects}}, is shown in Table \ref{table:dataSet}, \textit{ (e.g. apache{\tiny hadoop}\footnote{Apache {\tiny hadoop} \url{http://hadoop.apache.org/}}, apache{\tiny hive}\footnote{Apache {\tiny hive} \url{https://hive.apache.org/}}, apache{\tiny phoenix}\footnote{Apache {\tiny phoenix} \url{https://phoenix.apache.org/}}, apache{\tiny pig}\footnote{Apache {\tiny pig} \url{https://pig.apache.org/}}, apache{\tiny tomcat}\footnote{Apache {\tiny tomcat} \url{http://tomcat.apache.org/}}, apache{\tiny nutch}\footnote{Apache {\tiny nutch} \url{http://nutch.apache.org/}}, apache{\tiny ant core}\footnote{Apache {\tiny ant core} \url{http://ant.apache.org/}}, aspectJ{\tiny Aspect Oriented Frameworks}\footnote{aspectJ {\tiny Aspect Oriented Frameworks} \url{https://www.eclipse.org/aspectj/}}, jEdit{\tiny Programmer\'s Text Editor}\footnote{jEdit {\tiny Programmer\'s Text Editor} \url{http://www.jedit.org/}}, JFreeChart\footnote{JFreeChart \url{http://www.jfree.org/jfreechart/}}, JHotDraw\footnote{JHotDraw \url{http://www.jhotdraw.org/}}, JUnit4\footnote{JUnit4 \url{http://junit.org/junit4/}}, libgdx{\tiny Java game development framework}\footnote{libgdx {\tiny Java game development framework} \url{https://libgdx.badlogicgames.com/}}, openjms{\tiny Java Message Service}\footnote{openjms {\tiny Java Message Service} \url{http://openjms.sourceforge.net/}}, and scarab{\tiny Issue Tracking}\footnote{scarab {\tiny Issue Tracking} \url{https://java-source.net/open-source/issue-trackers/scarab}} )}. 

\begin{table}[h]
\caption{Data set of 15 open source projects as input to DPVIA tool}
\label{table:dataSet}
\begin{tabular}{p{5cm} p{2cm} p{2cm} p{2.8cm}}

\hline  
 
Project name & Lines of Code & Source Files & Total Detected patterns\\
\hline

apache {\tiny hadoop} & 1214896 & 5519 & 1093\\
apache {\tiny hive} & 1034094 & 3766 & 838\\
apache {\tiny phoenix} & 222353 & 850 & 590\\
apache {\tiny pig} & 398403 & 1765 & 831\\
apache {\tiny tomcat} & 537724 & 2240 & 64\\ 
apache {\tiny nutch} & 81543 & 536 & 50\\
apache {\tiny ant core} & 267028 & 1233 & 481\\
aspectJ {\tiny Aspect Oriented Frameworks} & 710700 & 7048 & 522\\
jEdit{\tiny  Programmer\'s Text Editor} & 195952 & 598 & 41\\
JFreeChart & 297386 & 993 & 4045\\
JHotDraw 6& 73421 & 491 & 155\\
JUnit4 & 43073 & 443 & 26\\
libgdx {\tiny Java game development framework} & 384745 & 2163 & 175\\
openjms {\tiny Java Message Service} & 112410 & 576 & 297\\
scarab {\tiny Issue Tracking} & 106236 & 448 & 30\\

\hline 
\end{tabular} 
\end{table}

The DPVIA, as it's result is shown in Table \ref{table:result}, identified the conformance scores for 9,238 pattern instances of seven different GoF patterns: Adapter, Decorator, Factory Method, Simple Factory, Observer, State and Strategy. The similarity scores indicates the conformance for pattern candidates with pattern definitions characteristics for each project in the repository, we observed that \textbf{\textit{open source projects have some instances of design patterns do not have a conformance between pattern implementations and their predefined characteristics, and this may cause a lack of maintainability}}.


In addition, we observed that \textbf{\textit{the proposed approach is able to assess, validate violations, and recommend a suitable solutions for small and large scale project of Java applications}}, as shown in Table \ref{table:dataSet}, the DPVIA tool receives as one input 15 open source Java project with different size. For each project, pattern candidates are detected and measure the conformance score for all candidate members versus the predefined characteristics of GoF pattern definitions. We argue that \textbf{\textit{validation of design pattern instances should be done based on source code files directly}} by parsing source code to extract the syntax parse tree (AST) which can be used for deeper analysis of the source elements.
\begin{table}[h]
\caption{Similarity conformance scores reported by DPVIA tool}
\label{table:result}
\begin{tabular}{p{2.6cm}| r r r r r r r}

\multicolumn{8}{c}{GoF design patterns} \\
\hline  
 
Project name & Adapter & Decorator & FactoryM & SFactory & Observer & \hspace{2mm} State & Strategy\\
\hline
 
hadoop & 100\% & 99.1\% & 92.5\% & 87.9\% & 85.2\% & 100\% & 91.6\%\\
hive & 100\% & 90.5\% & 93.1\% & 84.7\% & 85\% & 100\% & 91.7\%\\
phoenix & 96.5\% & 83\% & 98.7\% & 99\% & 91.8\% & -  & - \\
pig & 96.1\% & 94.2\% & 87.2\% & 85\% & 100\% & 91.6\% & - \\
tomcat & 99.1\% & 85\% & 100\% & 91.5\% & -  & -  & - \\
nutch & 100\% & 85\% & 91.9\% & -  & -  & -  & - \\
ant- core & 97.2\% & 100\% & 83\% & 85\% & 91.7\% & -  & - \\
aspectJ & 100\% & 92.5\% & 91.8\% & 93.2\% & 87.2\% & 100\% & 91.7\%\\
jEdit & 100\% & 85.7\% & 100\% & 91.5\% & -  & -  & - \\
JFreeChart & 100\% & 94.8\% & 97.9\% & 85\% & 100\% & 91.5\% & - \\
jhotdraw6 & 100\% & 95\% & 88.4\% & 100\% & 91.9\% & -  & - \\
junit4 & 100\% & 91.5\% & 87.2\% & 92\% & -  & -  & - \\
libgdx & 100\% & 93.4\% & 93.8\% & 94\% & 86.2\% & 100\% & 91.5\%\\
openjmsJMS & 95\% & 91.5\% & 87\% & 100\% & 91.8\% & -  & - \\
scarab & 83\% & 90\% & 91.5\% & -  & -  & -  & - \\
\hline 
Average & 97.8\% &91.4\% &92.3\% &91.4\% &91.1\% &97.2\% &91.6\%  \\
\hline
\end{tabular} 
\end{table}

\textbf{\textit{DPVIA is fully customizable}} since it allows developers to configure the definition of the patterns structure and their behavior, as well the developers are able to specify the predefined characteristics of any pattern that used in assessment the pattern implementations.

\section{Conclusion}

The major contribution of this work, to the domain of design patterns, includes an approach for automated identification of design pattern violations occurring in different project implementations and recommend a suitable solutions for software developer to modify the detected pattern violations. The detection of design patterns violations is done by measuring the conformance scoring of the implemented design patterns towards their definitions characteristics. That's why we developed an automated tool named \textit{Design Pattern Violations Identification and Assessment (DPVIA)}, in order to detect design patterns occurring in different projects implementations, and measure the conformance score for each pattern candidate to identify its violations. In addition, DPVIA tool reports violation details with appropriate solution as recommendations based on predefined pattern characteristics, then visualizes the results in charts for indicating the percentage of violation that has been committed. The violation is committed after proving the existence of relationships between its members in business logic (SRS document), which is detected by the Stanford CoreNLP Natural Language Processing Toolkit \cite{StanfordNLP} to provide a valuable insight on design pattern violations assessment.

We sincerely hope that this work will inspire further researches in this field, for instance the detected pattern violations would be re-factored or discarded once identified, but that would added massive amount of work to developers in order to re-factor those violations. As well, the decision of applying the recommended solutions for the detected pattern violations is usually a trade-off, because patterns are not universally good or bad. Patterns typically improve certain aspects of software quality, while they might weaken some other. For these reasons we look forward to build a semi-automated violations re-factoring module to fix detected violations in Java project source code. Finally, according to the efficient execution time and minimum misleading pattern violations identification, we believe the proposed DPVIA tool is an efficient alternative to existing tools.



\bibliographystyle{ieeetr}
\bibliography{refs}

\begin{thebibliography}{10}

\bibitem{gof}
R.~J. Erich~Gamma, Richard~Helm and J.~Vlissides, {\em Design Patterns:
  Elements of reusable object-oriented software}.
\newblock Addison-Wesley, 1995.

\bibitem{designroles}
Y.-G.~G. Foutse~Khomh and G.~Antonio, ``Playing roles in design patterns: An
  empirical descriptive and analytic study,'' {\em In: 25th IEEE International
  Conference on Software Maintenance. IEEE}, pp.~83--92, 2009.

\bibitem{designroles2}
S.~C. Apostolos~Ampatzoglou, Alexander~Chatzigeorgiou and P.~Avgeriou, ``The
  effect of gof design patterns on stability: A case study,'' {\em I IEEE
  Trans. Softw. Eng. 41, 781–802}, 2015.

\bibitem{Riehle}
D.~Riehle, ``Lessons learned from using design patterns in industry
  projects.,'' {\em In Transactions on Pattern Languages of Programming II,
  Springer-Verlag}, vol.~LNCS 6510, pp.~1--15, 2011.

\bibitem{Bautista}
N.~Bautista, ``A beginners guide to design patterns.,'' {\em Accessed August
  15, 2017.}, \url{http://code.tutsplus.com/articles/a-
  beginners-guide-to-design-patterns--net-12752}.

\bibitem{Ampatzoglou}
S.~C. Apostolos~Ampatzoglou and I.~Stamelos, ``Research state of the art on gof
  design patterns: A mapping study.,'' {\em Journal of Systems and Software,
  Elsevier}, vol.~86, no.~7, pp.~1945--1964, July 2013.

\bibitem{Ampatzoglou2}
S.~C. Apostolos~Ampatzoglou and I.~Stamelos, ``Design pattern alternatives:
  What to do when a gof pattern fails.,'' {\em Proceedings of the 17th
  Panhellenic Conference on Informatics At: Thessaloniki, Greece}, pp.~1--6,
  September 2013.

\bibitem{Iyad}
B.~A.-H. Iyad~Alazzam and E.~Migdady, ``Design patterns detection based on its
  domain.,'' {\em Information Technology (ICIT) 2017 8th International
  Conference}, pp.~304--308, 2017.

\bibitem{Parnas}
D.~L. Parnas, ``Software aging.,'' {\em ICSE '94 Proceedings of the 16th
  international conference on Software engineering, IEEE Computer Society Press
  Los Alamitos, CA, USA}, pp.~279--287, 1994.

\bibitem{J.M.Bieman}
H.~W. P. W.~M. J.~M.~Bieman, G.~Straw and R.~T. Alexander, ``Design patterns
  and change proneness: an examination of five evolving systems,'' {\em
  Proceedings. 5th International Workshop on Enterprise Networking and
  Computing in Healthcare Industry (IEEE Cat. No.03EX717)}, pp.~40--49, 2003.

\bibitem{M.Gatrell}
S.~C. M.~Gatrell and T.~Hall, ``Design patterns and change proneness: A
  replication using proprietary c\# software,'' {\em 2009 16th Working
  Conference on Reverse Engineering, Lille}, pp.~160--164, 2009.

\bibitem{Tsantalis}
G.~S. Nikolaos~Tsantalis, Alexander~Chatzigeorgiou and S.~T. Halkidis, ``Design
  pattern detection using similarity scoring,'' {\em IEEE Transactions on
  Software Engineering}, vol.~32, no.~11, pp.~896--909, 2006.

\bibitem{GraphAlgo}
M.~H. P.~S. Vincent D.~Blondel, Anahi~Gajardo and P.~V. Dooren, ``A measure of
  similarity between graph vertices: Applications to synonym extraction and web
  searching,'' {\em SIAM Rev.}, vol.~46, no.~4, pp.~647--666, 2004.

\bibitem{Izurieta}
C.~Izurieta and J.~M. Bieman, ``How software designs decay: A pilot study of
  pattern evolution,'' {\em First International Symposium on Empirical Software
  Engineering and Measurement}, pp.~ESEM 459--461, 2007.

\bibitem{Izurieta2}
C.~Izurieta, ``Decay and grime buildup in evolving object oriented design
  patterns,'' {\em Colorado State University Fort Collins}, 2009.

\bibitem{Izurieta1}
C.~Izurieta and J.~M.Bieman, ``A multiple case study of design pattern decay,
  grime, and rot in evolving software systems,'' {\em in Software Quality
  Journal (2013) Springer Science+ Business Media}, pp.~289--323, 2012.

\bibitem{Moha}
D.-l.~H. Naouel~Moha and Y.-G. Gueheneuc, ``A taxonomy and a first study of
  design pattern defects,'' {\em IEEE International Workshop on Software
  Technology and Engineering Practice, IEEE Computer Society, Budapest,
  Hungary}, pp.~225--229, 2005.

\bibitem{Dale}
M.~R. Dale and C.~Izurieta, ``Impacts of design pattern decay on system
  quality,'' {\em ESEM 14 Proceedings of the 8th ACM/IEEE International
  Symposium on Empirical Software Engineering and Measurement, ACM Press, New
  York, NY, USA}, 2014.

\bibitem{Strasser}
K.~F. Shane~Strasser, Colt~Frederickson and C.~Izurieta, ``An automated
  software tool for validating design patterns,'' {\em Honolulu}, 2011.

\bibitem{Kim}
S.~G. Dae-Kyoo~Kim, Robert~France and E.~Song, ``Using role-based modeling
  language (rbml) to characterize model families,'' {\em In Eighth IEEE
  International Conference on Engineering of Complex Computer Systems}, 2002.

\bibitem{Diamantopoulos}
A.~N. Themistoklis~Diamantopoulos and A.~Symeonidis, ``Dp-core: A design
  pattern detection tool for code reuse,'' {\em Proceedings of the Sixth
  International Symposium on Business Modeling and Software Design (BMSD)},
  pp.~160--169, 2016.

\bibitem{StanfordNLP}
M.~S. J. B. J. F. S. J.~B. Manning, Christopher~D. and D.~McClosky., ``The
  stanford corenlp natural language processing toolkit,'' {\em in Proceedings
  of the 52nd Annual Meeting of the Association for Computational Linguistics:
  System Demonstrations}, pp.~55--60, 2014.

\bibitem{Hamming}
R.~W. Hamming, ``Error detecting and error correcting codes,'' {\em The Bell
  System Technical Journal}, vol.~29, no.~2, pp.~147--160, April 1950.

\bibitem{StanfordOperIE}
M.~J.~P. Gabor~Angeli and C.~D. Manning., ``Leveraging linguistic structure for
  open domain information extraction,'' {\em In Proceedings of the Association
  of Computational Linguistics (ACL)}, 2015.

\end{thebibliography}

\end{document}